\documentclass[twocolumn,superscriptaddress,
showpacs,preprintnumbers,amsmath,amssymb]{revtex4}
\usepackage{graphicx}% Include figure files

\newcommand{\bra}[1]{{\left\langle #1 \right|}}
\newcommand{\ket}[1]{{\left| #1 \right\rangle}}

\begin{document}

%%%%%%%%%%%%%%%%%%%%%%%%%%%%%%%%%%%%%%%%%%%%%%%%%%%%%%%%%%%%%%%%%%%%%%%%%%
%                                                                        %
%                                 Title                                  %
%                                                                        %
%%%%%%%%%%%%%%%%%%%%%%%%%%%%%%%%%%%%%%%%%%%%%%%%%%%%%%%%%%%%%%%%%%%%%%%%%%
\title{Entanglement of three-qubit pure states in terms of teleportation capability}

\author{Soojoon Lee}\email{level@khu.ac.kr}
\affiliation{
 Department of Mathematics and Research Institute for Basic Sciences,
 Kyung Hee University, Seoul 130-701, Korea
}
\author{Jaewoo Joo}\email{jaewoo.joo@imperial.ac.uk}
\affiliation{
 Blackett Laboratory, Imperial College London,
 Prince Consort Road, London, SW7 2BW, UK
}
\author{Jaewan Kim}\email{jaewan@kias.re.kr}
\affiliation{
 School of Computational Sciences,
 Korea Institute for Advanced Study,
 Seoul 130-722, Korea
}
\date{\today}

%%%%%%%%%%%%%%%%%%%%%%%%%%%%%%%%%%%%%%%%%%%%%%%%%%%%%%%%%%%%%%%%%%%%%%%%%%
%                                                                        %
%                              Abstract                                  %
%                                                                        %
%%%%%%%%%%%%%%%%%%%%%%%%%%%%%%%%%%%%%%%%%%%%%%%%%%%%%%%%%%%%%%%%%%%%%%%%%%
\begin{abstract}
We define an entanglement measure, called the partial tangle,
which represents the residual two-qubit entanglement of a three-qubit pure state.
By its explicit calculations for three-qubit pure states,
we show that the partial tangle is closely related to
the faithfulness of a teleportation scheme over a three-qubit pure state.
\end{abstract}

%\keywords{}
\maketitle

%%%%%%%%%%%%%%%%%%%%%%%%%%%%%%%%%%%%%%%%%%%%%%%%%%%%%%%%%%%%%%%%%%%%%%
%%%                                                                %%%
%%%                         Introduction                           %%%
%%%                                                                %%%
%%%%%%%%%%%%%%%%%%%%%%%%%%%%%%%%%%%%%%%%%%%%%%%%%%%%%%%%%%%%%%%%%%%%%%
%%%
%%%     Teleportation and entanglement
%%%
Quantum entanglement has been considered to be one of the most crucial resources
in quantum information processing, and hence
has been studied intensively in various ways.
Nevertheless, there are still a number of open problems for entanglement,
such as
what is the best way to
quantify the amount of entanglement for bipartite or multipartite states.

For two-qubit states,
the Wootters' concurrence $\mathcal{C}$~\cite{BDSW,HW,Wootters},
is known as a good measure of entanglement,
since from it we can directly derive the explicit formula for the entanglement of formation
as well as being readily calculable.
On the other hand, in the multi-qubit cases, or even in the three-qubit case,
no entanglement measure as good as the concurrence of two qubits
has been found yet.

Coffman {\it et al.}~\cite{CKW} presented
an inequality to explain the relation between bipartite entanglement
in a three-qubit pure state.
The inequality is called
the Coffman-Kundu-Wootters (CKW) inequality,
which is
\begin{equation}
\mathcal{C}^2_{12}+\mathcal{C}^2_{13}\le \mathcal{C}^2_{1(23)},
\label{eq:CKW3}
\end{equation}
where $\mathcal{C}_{12}=\mathcal{C}(\mathrm{tr}_3(\Psi_{123}))$,
$\mathcal{C}_{13}=\mathcal{C}(\mathrm{tr}_2(\Psi_{123}))$,
and $\mathcal{C}_{1(23)}=\mathcal{C}(\Psi_{1(23)})=2\sqrt{\det(\mathrm{tr}_{23}(\Psi_{123}))}$
for a three-qubit pure state $\Psi_{123}=\ket{\psi}_{123}\bra{\psi}$.
Here, the subscripts represent the indices of the qubits.

From the CKW inequality,
an entanglement measure for three-qubit pure states
was naturally derived~\cite{CKW,DVC}.
It is called the 3-tangle $\tau$, which is defined as
\begin{equation}
\tau=\mathcal{C}^2_{1(23)}-\mathcal{C}^2_{12}-\mathcal{C}^2_{13},
\label{eq:tangle}
\end{equation}
and represents the residual entanglement of the state.
Here $\tau$ is invariant under any qubit taken as the focus qubit,
that is, for any distinct $i$, $j$, and $k$ in $\{1,2,3\}$,
\begin{equation}
\tau=\mathcal{C}^2_{i(jk)}-\mathcal{C}^2_{ij}-\mathcal{C}^2_{ik}.
\label{eq:tau_invariance}
\end{equation}
Furthermore, it was shown that $\tau$ is an entanglement monotone~\cite{DVC},
and it was also shown that $\tau$ can distinguish
the Greenberger-Horne-Zeilinger (GHZ) class
from the W class~\cite{DVC},
where the GHZ class and the W class are the sets of all pure states with true three-qubit entanglement
equivalent to the GHZ state~\cite{GHZ},
\begin{equation}
\ket{GHZ}=\frac{1}{\sqrt{2}}\left(\ket{000}+\ket{111}\right),
\label{eq:GHZ}
\end{equation}
under stochastic local operations and classical communication (SLOCC),
and equivalent to the W state,
\begin{equation}
\ket{W}=\frac{1}{\sqrt{3}}\left(\ket{001}+\ket{010}+\ket{100}\right),
\label{eq:W}
\end{equation}
under SLOCC, respectively.

Even though the 3-tangle $\tau$ is a useful entanglement measure for three-qubit pure states,
in this paper,
we investigate another quantity similar to $\tau$, %on three-qubit pure states,
defined as
\begin{equation}
\tau_{ij}=\sqrt{\mathcal{C}^2_{i(jk)}-\mathcal{C}^2_{ik}},
\label{eq:def_partial_tangle}
\end{equation}
for distinct $i$, $j$, and $k$ in $\{1,2,3\}$.
We call the quantity the {\it partial tangle}.
Then we clearly obtain the following equalities:
\begin{eqnarray}
\tau_{12}&=&\sqrt{\mathcal{C}^2_{1(23)}-\mathcal{C}^2_{13}}=\sqrt{\tau+\mathcal{C}^2_{12}}=\tau_{21},
\nonumber\\
\tau_{23}&=&\sqrt{\mathcal{C}^2_{2(31)}-\mathcal{C}^2_{21}}=\sqrt{\tau+\mathcal{C}^2_{23}}=\tau_{32},
\nonumber\\
\tau_{31}&=&\sqrt{\mathcal{C}^2_{3(12)}-\mathcal{C}^2_{32}}=\sqrt{\tau+\mathcal{C}^2_{31}}=\tau_{13},
\label{eq:tau3}
\end{eqnarray}
and hence
\begin{equation}
\tau_{12}^2+\tau_{23}^2+\tau_{31}^2=3\tau + \mathcal{C}^2_{12}+\mathcal{C}^2_{23}+\mathcal{C}^2_{31}.
\label{eq:tau_concurrence}
\end{equation}
We clearly remark that $\tau_{ij}=\mathcal{C}_{ij}$
if and only if a given state is contained in the W class, that is, $\tau=0$.

Observing the definition of $\tau_{ij}$ in Eq.~(\ref{eq:tau3}),
%each of them
$\tau_{ij}$ seems to represent the residual two-qubit entanglement of a three-qubit pure state.
However, we cannot say that
$\tau_{ij}$ represents only the entanglement for two qubits in the compound system $ij$
since $\tau_{ij}$ is not equivalent to $\mathcal{C}_{ij}$ in general as in Eq.~(\ref{eq:tau3}).
Therefore, in order to understand the entanglement of three-qubit states more evidently,
it would be important to investigate the meaning of $\tau_{ij}$.

In this paper,
we explicitly calculate the partial tangle for three-qubit pure states
so as to investigate its meaning,
and we show that the partial tangle is closely related to a teleportation scheme over three-qubit pure states
as a relation between the concurrence and the fully entangled fraction for two-qubit pure states.

We note that any three-qubit pure state $\ket{\psi}$ can be written in the form~\cite{AACJLT,ABLS}
\begin{eqnarray}
\ket{\psi}&=&\lambda_0\ket{000}+\lambda_1 e^{\iota\theta}\ket{100}+\lambda_{2}\ket{101}\nonumber\\
&&+\lambda_{3}\ket{110}+\lambda_{4}\ket{111},
\label{eq:Schmidt3}
\end{eqnarray}
where $\iota=\sqrt{-1}$, $0\le\theta\le\pi$, $\lambda_j\ge 0$, and $\sum_j\lambda_j^2=1$.
Thus, in order to calculate the partial tangles for three-qubit pure states,
it suffices to consider the ones for the states in Eq.~(\ref{eq:Schmidt3}).
By somewhat tedious but straightforward calculations,
we obtain the following results on the partial tangles $\tau_{ij}$ for $\ket{\psi}$:
\begin{eqnarray}
\tau_{12}&=&2\lambda_0\sqrt{\lambda^2_3+\lambda^2_4},\nonumber\\
\tau_{23}&=&2\sqrt{\lambda^2_0\lambda^2_4+\lambda^2_1\lambda^2_4+\lambda^2_2\lambda^2_3
-2\lambda_1\lambda_2\lambda_3\lambda_4\cos\theta},\nonumber\\
\tau_{31}&=&2\lambda_0\sqrt{\lambda^2_2+\lambda^2_4}.
\label{eq:calculation_tau}
\end{eqnarray}

Since one of the most important practical features of entanglement
is the teleportation capability,
we now consider a teleportation scheme over a three-qubit state in the compound system $123$,
which is a modification of the splitting and reconstruction of quantum information over the GHZ state,
introduced by Hillery {\it et al.}~\cite{HBB}.
\begin{figure}
\includegraphics[angle=-90,scale=0.90,width=\linewidth]{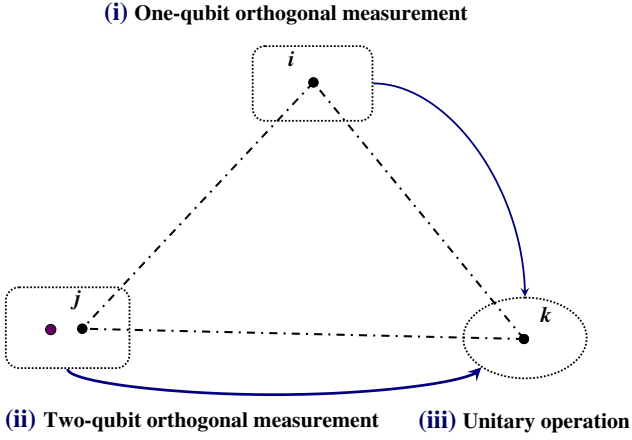}
\caption{\label{Fig:diagram}
Our modified teleportation scheme over a three-qubit state:
The dotted boxes and ellipse represent performing the orthogonal measurements
and applying the unitary operation, respectively.
The arrows represent sending classical information corresponding to the measurement results.
}
\end{figure}
The modified scheme is illustrated in Fig.~\ref{Fig:diagram}
and is described as follows:
Let $i$, $j$, and $k$ be distinct in $\{1,2,3\}$.
(i)~Make a one-qubit orthogonal measurement on the system $i$.
(ii)~Prepare an arbitrary one-qubit state,
and then make a two-qubit orthogonal measurement on the one qubit and the system $j$.
(iii)~On the system $k$,
apply a proper unitary operation
related to the 3-bit classical information of the two above measurement results.

We note that this scheme is nothing but a teleportation over the two-qubit state on the systems $j$ and $k$
after the measurement of the system~$i$,
and that the faithfulness of this teleportation
completely depends on the probabilities corresponding to the one-qubit measurement results in step~(i)
and the resulting state of the systems $j$ and $k$ after the one-qubit measurement.

We remark that any observable for a one-qubit measurement can be described as
\begin{equation}
U^{\dagger}\sigma_3U=U^{\dagger}\ket{0}\bra{0}U-U^{\dagger}\ket{1}\bra{1}U,
\label{eq:observable}
\end{equation}
where $\sigma_3=\ket{0}\bra{0}-\ket{1}\bra{1}$ is one of Pauli matrices,
and $U$ is a $2\times 2$ unitary matrix.
Thus, after the step~(i) of the teleportation scheme over $\ket{\psi}$,
the resulting 2-qubit state of the compound system $jk$ becomes
\begin{eqnarray}
\varrho_{jk}^{t}
&\equiv& \frac{\mathrm{tr}_{i}
\left(U_i^{\dagger}\ket{t}\bra{t}U_i\otimes I_{jk}
\ket{\psi}\bra{\psi}
U_i^{\dagger}\ket{t}\bra{t}U_i\otimes I_{jk}\right)}
{\bra{t}U_i\rho_{i}U_i^{\dagger}\ket{t}}\nonumber\\
&=& \frac{\mathrm{tr}_{i}
\left(\ket{t}\bra{t}U_i\otimes I_{jk}
\ket{\psi}\bra{\psi}
U_i^{\dagger}\ket{t}\bra{t}\otimes I_{jk}\right)}
{\bra{t}U_i\rho_{i}U_i^{\dagger}\ket{t}}
\end{eqnarray}
with probability $\bra{t}U_i\rho_{i}U_i^{\dagger}\ket{t}$
for $t=0$ or $1$,
where $U_i$ is a $2\times 2$ unitary matrix of the system~$i$,
and $\rho_i=\mathrm{tr}_{jk}(\ket{\psi}\bra{\psi})$.
Since $\varrho_{jk}^{t}$ is the resulting state after the orthogonal measurement,
it must be a 2-qubit pure state.
For example, if $i=1$, $j=2$, $k=3$, and
\begin{equation}
U_1=\left(
\begin{array}{cc}
u_{00} & u_{01} \\
u_{10} & u_{11}
\end{array}
\right)\in \mathrm{U}(2),
\label{eq:U}
\end{equation}
then
\begin{equation}
\varrho_{jk}^{t}\bra{t}U_i\rho_{i}U_i^{\dagger}\ket{t}=\ket{\psi_{jk}^t}\bra{\psi_{jk}^t},
\label{eq:varrhojk}
\end{equation}
where
\begin{eqnarray}
\ket{\psi_{jk}^t}
&=&(\lambda_0u_{0t}+\lambda_1e^{\iota\theta}u_{1t})\ket{00}\nonumber\\
&&+\lambda_2u_{1t}\ket{01}+\lambda_3u_{1t}\ket{10}+\lambda_4u_{1t}\ket{11}.
\end{eqnarray}

For the moment,
we shall review the properties of the faithfulness of a teleportation over a 2-qubit state.
This faithfulness %of a teleportation over a 2-qubit state $\rho$
is naturally provided by teleportation's fidelity~\cite{Popescu},
\begin{equation}
F(\Lambda_{\rho})=\int d\xi \bra{\xi}\Lambda_{\rho}(\ket{\xi}\bra{\xi})\ket{\xi},
\label{eq:teleportation_fidelity}
\end{equation}
where $\Lambda_\rho$ is a given teleportation scheme over a 2-qubit state $\rho$,
and the integral is performed
with respect to the uniform distribution $d\xi$ over all one-qubit pure states.
We also consider
the {\it fully entangled fraction}~\cite{BDSW,Horodeckis1,Horodeckis2,BadziagHorodeckis} of $\rho$
defined as
\begin{equation}
f(\rho)=\max\bra{e}\rho\ket{e},
\label{eq:FEF}
\end{equation}
where the maximum is over all maximally entangled states $\ket{e}$ of 2 qubits.
It has been shown~\cite{Horodeckis2,BadziagHorodeckis} that
the maximal fidelity achievable from a given bipartite state $\rho$ is
\begin{equation}
F(\Lambda_{\rho})=\frac{2f(\rho)+1}{3},
\label{eq:2relation}
\end{equation}
where $\Lambda_{\rho}$ is the standard teleportation scheme over $\rho$ to provide the maximal fidelity.
Furthermore,
for any two-qubit pure state $\ket{\phi}=\sqrt{\alpha}\ket{00}+\sqrt{\beta}\ket{11}$
with $\alpha$, $\beta\ge 0$ satisfying $\alpha+\beta=1$,
we can readily obtain that
\begin{eqnarray}
f(\ket{\phi}\bra{\phi})&=&1/2+\sqrt{\alpha\beta},\nonumber\\
\mathcal{C}(\ket{\phi}\bra{\phi})&=&2\sqrt{\alpha\beta},
\label{eq:f_C}
\end{eqnarray}
and hence
\begin{equation}
\mathcal{C}(\ket{\phi}\bra{\phi})=2f(\ket{\phi}\bra{\phi})-1
=3F(\Lambda_{\ket{\phi}\bra{\phi}})-2
\label{eq:C_f}
\end{equation}
for any two-qubit pure state $\ket{\phi}$.

%%%
%%%     Tripartite cases
%%%
Let us define $F_i$ as
the maximal teleportation's fidelity over the resulting 2-qubit state in the systems~$j$ and $k$
after the measurement of the system~$i$.
Then, from the above review,
it is straightforward to obtain that for $i\in \{1,2,3\}$
\begin{equation}
F_i=\frac{2f_i+1}{3},
\label{eq:Fi}
\end{equation}
where
\begin{equation}
f_i=\max_{U_i}\left[\bra{0}U_i\rho_{i}U_i^{\dagger}\ket{0}f(\varrho_{jk}^{0})
+\bra{1}U_i\rho_{i}U_i^{\dagger}\ket{1}f(\varrho_{jk}^{1})\right].
\label{eq:fi}
\end{equation}
Here, the maximum is over all $2\times 2$ unitary matrices.
Since $\varrho_{jk}^{t}$ is pure, $f_i$ can be rewritten as
\begin{align}
f_i=\frac{1}{2}\max_{U_i}
&\left[\bra{0}U_i\rho_{i}U_i^{\dagger}\ket{0}\left(1+\mathcal{C}(\varrho_{jk}^{0})\right)\right.
\nonumber\\
&\left.+\bra{1}U_i\rho_{i}U_i^{\dagger}\ket{1}\left(1+\mathcal{C}(\varrho_{jk}^{1})\right)\right].
\label{eq:fi2}
\end{align}
After tedious calculations~\cite{f1}, we get the following results:
\begin{eqnarray}
f_1&=&\frac{1}{2}+\sqrt{\lambda^2_0\lambda^2_4+\lambda^2_1\lambda^2_4+\lambda^2_2\lambda^2_3
-2\lambda_1\lambda_2\lambda_3\lambda_4\cos\theta},\nonumber\\
f_2&=&\frac{1}{2}+\lambda_0\sqrt{\lambda^2_2+\lambda^2_4},\nonumber\\
f_3&=&\frac{1}{2}+\lambda_0\sqrt{\lambda^2_3+\lambda^2_4}.
\label{eq:fis}
\end{eqnarray}
Therefore, it follows from Eqs.~(\ref{eq:calculation_tau}), (\ref{eq:Fi}) and~(\ref{eq:fis}) that
\begin{equation}
\tau_{ij}=2f_k-1=3F_k-2.
\label{eq:main_relation}
\end{equation}
We remark that $f_i\ge 1/2$ and $F_i\ge 2/3$ for three-qubit pure states,
and that the above result in Eq. (\ref{eq:main_relation})
is surprisingly of the same form as that in Eq.~(\ref{eq:C_f}).
Thus, we could say that $\tau_{ij}$ is a three-qubit version of the concurrence
with respect to a teleportation over a three-qubit pure state.
Moreover, it could be meaningful that
a kind of mathematical quantity, $\tau_{ij}$, is closely concerned with
$f_k$ and $F_k$ as the quantities derived from physical information processing,
as in the two-qubit case.

%%%%%%%%%%%%%%%%%%%%%%%%%%%%%%%%%%%%%%%%%%%%%%%%%%%%%%%%%%%%%%%%%%%%%%
%%%                                                                %%%
%%%                         Conclusion                             %%%
%%%                                                                %%%
%%%%%%%%%%%%%%%%%%%%%%%%%%%%%%%%%%%%%%%%%%%%%%%%%%%%%%%%%%%%%%%%%%%%%%
In conclusion,
we have considered the so-called partial tangle $\tau_{ij}$
as an entanglement measure for three-qubit pure states. %derived from these inequalities.
We have also considered the quantities $f_k$ and $F_k$ obtained from
the maximal fidelity of a teleportation scheme over a three-qubit pure state.
By their explicit calculations for three-qubit pure states,
we have shown that there exists a close relation between
the mathematical quantity $\tau_{ij}$ related to the three-qubit entanglement
and the physical quantities $f_k$ and $F_k$ related to the teleportation capability,
as in the two-qubit case.

%%%%%%%%%%%%%%%%%%%%%%%%%%%%%%%%%%%%%%%%%%%%%%%%%%%%%%%%%%%%%%%%%%%%%%
%%%                                                                %%%
%%%                       Acknowledgements                         %%%
%%%                                                                %%%
%%%%%%%%%%%%%%%%%%%%%%%%%%%%%%%%%%%%%%%%%%%%%%%%%%%%%%%%%%%%%%%%%%%%%%
%\acknowledgments{
S.L. acknowledges V. Bu\v{z}ek and M. Horodecki for encouraging discussions,
and J.J. thanks M.B. Plenio for useful advices.
J.J. was supported by the Overseas Research Student Award Program for financial support,
and J.K. by a Korea Research Foundation Grant (KRF-2002-070-C00029).
%}


\begin{thebibliography}{1}
\bibitem{f1}
For example, we consider the case that $i=1$, $j=2$, $k=3$, and
$U_1$ is the same as that in Eq.~(\ref{eq:U}), and hence
$\varrho_{23}^t$ is the same as that in Eq.~(\ref{eq:varrhojk}) for $t=0$ or 1.
Since $\mathcal{C}(\varrho_{23}^t)=2\sqrt{\det(\mathrm{tr}_3(\varrho_{23}^t))}$ and
$\sum_{t}\bra{t}U_i\rho_{i}U_i^{\dagger}\ket{t}=1$,
we obtain
\begin{eqnarray}
f_1&=&\frac{1}{2}+\max_{U_1}
\sum_{t=0}^1 \left|u_{1t}\right|\left|\lambda_4\left(\lambda_0 u_{0t}+\lambda_1 u_{1t}e^{\iota\theta}\right)
-\lambda_2\lambda_3\lambda_4 u_{1t}\right|\nonumber\\
&\le&
\frac{1}{2}+\max_{U_1}
\sqrt{
\sum_{t=0}^1 \left|\lambda_4\left(\lambda_0 u_{0t}+\lambda_1 u_{1t}e^{\iota\theta}\right)
-\lambda_2\lambda_3\lambda_4 u_{1t}\right|^2}\nonumber\\
&=&\frac{1}{2}+\sqrt{\lambda^2_0\lambda^2_4+\lambda^2_1\lambda^2_4+\lambda^2_2\lambda^2_3
-2\lambda_1\lambda_2\lambda_3\lambda_4\cos\theta},
\end{eqnarray}
where the inequality is a consequence of the Cauchy-Schwarz inequality,
and the last equality is independent of the above maximum,
that is,
\begin{align}
\sum_{t=0}^1 & \left|\lambda_4\left(\lambda_0 u_{0t}+\lambda_1 u_{1t}e^{\iota\theta}\right)
-\lambda_2\lambda_3\lambda_4 u_{1t}\right|^2\nonumber\\
&= \lambda^2_0\lambda^2_4+\lambda^2_1\lambda^2_4+\lambda^2_2\lambda^2_3
-2\lambda_1\lambda_2\lambda_3\lambda_4\cos\theta,
\end{align}
for any $2 \times 2$ unitary $U_1$.
Since we can readily check that there exists a $2 \times 2$ unitary matrix
providing the equality of the Cauchy-Schwarz inequality,
we can prove the first equality in Eq.~(\ref{eq:fis}).
The other equalities in Eq.~(\ref{eq:fis}) can be obtained in the same way.
\end{thebibliography}
\end{document}